# Sensitivity Assessing to Data Volume for forecasting: introducing similarity methods as a suitable one in Feature selection methods.

Mahdi Goldani

Soraya Asadi Tirvan

## Abstract

In predictive modeling, overfitting poses a significant risk, particularly when the feature count surpasses the number of observations, a common scenario in high-dimensional data sets. To mitigate this risk, feature selection is employed to enhance model generalizability by reducing the dimensionality of the data. This study focuses on evaluating the stability of feature selection techniques with respect to varying data volumes, particularly employing time series similarity methods. Utilizing a comprehensive dataset that includes the closing, opening, high, and low prices of stocks from 100 high-income companies listed in the Fortune Global 500, this research compares several feature selection methods including variance thresholds, edit distance, and Hausdorff distance metrics. The aim is to identify methods that show minimal sensitivity to the quantity of data, ensuring robustness and reliability in predictions, which is crucial for financial forecasting. Results indicate that among the tested feature selection strategies, the variance method, edit distance, and Hausdorff methods exhibit the least sensitivity to changes in data volume. These methods therefore provide a dependable approach to reducing feature space without significantly compromising the predictive accuracy. This study not only highlights the effectiveness of time series similarity methods in feature selection but also underlines their potential in applications involving fluctuating datasets, such as financial markets or dynamic economic conditions. The findings advocate for their use as principal methods for robust feature selection in predictive analytics frameworks.

### Introduce

In machine learning models, having a complete and comprehensive dataset can significantly enhance model accuracy. However, there are instances where the inclusion of irrelevant features may actually hinder rather than help the model's performance. In fact, the feature space with larger dimensions creates a larger number of parameters that need to be estimated. As a result, by increasing the number of parameters, the probability of overfitting in the demand model is strengthened. Therefore, the best performance generalization is achieved when a subset of the available features is used. Dimensionality reduction is the way to solve this challenge. The literature on dimensionality reduction refers to transforming data from a high-dimensional space into a low dimensional space. One of the most well-known techniques of dimensionality reduction is Feature selection [1]. Feature selection selects a subset of relevant features for use in model construction. The filters, embedded, and wrapper methods are the three main categories of Feature selection methods. Filter methods are characterized by their independence from specific machine learning algorithms. They prioritize data relationships, making them computationally efficient and straightforward to implement. In contrast, wrappers and embedded methods rely on learning algorithms. While filters are computationally efficient and easy to implement, wrappers often achieve better performance by considering feature interactions, albeit with increased computational complexity. Embedded methods strike a balance between filters and wrappers, integrating feature selection into the training process. This integration reduces computational costs compared to wrappers, as it eliminates the need for separate iterative evaluation of feature subsets [2]. Along with feature selection methods, time series similarity methods used for clustering in machine learning can be a suitable option for selecting a suitable subset of variables. similarity methods can serve as effective feature selection techniques by identifying redundant or irrelevant features, grouping similar features together, and quantifying the relationships between features and the target variable. By leveraging similarity measures in feature selection, one can extract the most informative features from the dataset while reducing dimensionality and improving model performance. The review of articles in this field shows that similarity methods are used in combination with feature selection methods.

In the literature, the primary aim of feature selection is to eliminate irrelevant variables, particularly when the number of features exceeds the number of observations. This practice helps mitigate overfitting, ensuring that the model generalizes well to unseen data. Therefore, feature selection is a method for dealing with a small number of observations. But does the performance of feature selection methods change when the number of observations is very small? In fact, this article seeks to find the answer to this question; When we are faced with a small number of observations, the results of which of the feature selection methods can be more reliable? This issue is important because most of the existing datasets that provide annual data face the problem of a small number of observations. Therefore, finding a way to reduce the dimension of a data set when the number of observations is low can help to increase the accuracy of the models. The aim of this research is to find the most optimal method to reduce the dimension of data that has the least impact on the performance of these models.

Feature selection is a widely used technique in various data mining and machine learning applications. In the literature on feature selection, there is no study that uses similarity methods directly as feature selection methods but there are some researches that explore this concept or incorporate similarity measures into feature selection processes. For example, Zhu et al [3] In the proposed Feature Selection-based Feature Clustering (FSFC) algorithm, similarity-based feature clustering utilized a means of unsupervised feature selection. Mitra [4] proposes an unsupervised feature selection algorithm designed for large datasets with high dimensionality. The algorithm is focused on measuring the similarity between features to identify and remove redundancy, resulting in a more efficient and effective feature selection process. In the domain of software defect prediction, Yu et al. [5] emphasize the central role of similarity in gauging the likeness or proximity among distinct software modules (referred to as samples) based on their respective features. Shi et al. [6] proposed a novel approach called Adaptive-Similarity-based Multi-modality Feature Selection (ASMFS) for multimodal classification in Alzheimer's disease (AD) and its prodromal stage, mild cognitive impairment (MCI). They addressed the limitations of traditional methods, which often rely on pre-defined similarity matrices to depict data structure, making it challenging to accurately capture the intrinsic relationships across different modalities in high-dimensional space. In the FU's [7] article Following the evaluation of feature relevance, redundant features are identified and removed using feature similarity. Features that exhibit high similarity to one another are considered redundant and are consequently eliminated from the dataset. Feature similarity measures are utilized to quantify the similarity between pairs of features. These measures help identify redundant features by assessing their degree of resemblance or closeness.

In terms of data size, there's been a bunch of studies that have addressed this issue. Vabalas [8] highlights the crucial role of sample size in machine learning studies, particularly in predicting autism spectrum disorder from high-dimensional datasets. It discusses how small sample sizes can lead to biased performance estimates and investigates whether this bias is due to validation methods not adequately controlling overfitting. Simulations show that certain validation methods produce biased estimates, while others remain robust regardless of sample size. Perry et al. [9] underscore the significance of sample size in machine learning for predicting geomorphic disturbances, showing that small samples can yield effective models, especially for identifying key predictors. It emphasizes the importance of thoughtful sampling strategies, suggesting that careful consideration can enhance predictive performance even with limited data. Cui and Goan [10] tested Six common ML regression algorithms on resting-state functional MRI (rs-fMRI) data from the Human Connectome Project (HCP), using various sample sizes ranging from 20 to 700. Across algorithms and feature types, prediction accuracy and stability increase exponentially with larger sample sizes. Kuncheva et al. [11] conducted experiments on 20 real datasets. In an exaggerated scenario, where only a small portion of the data (10 instances per class) was used for feature selection while the rest was reserved for testing, the results underscore the caution needed when performing feature selection on wide datasets. The findings suggest that in such cases, it may be preferable to avoid feature selection altogether rather than risk providing misleading results to users. Kuncheva [12] challenges the traditional feature selection protocol for high-dimensional datasets with few instances, finding it leads to biased accuracy estimates. It proposes an alternative protocol integrating feature selection and classifier testing within a single cross-validation loop, which yields significantly closer agreement with true accuracy estimates. This highlights the importance of re-evaluating standard protocols for accurate performance evaluation in such datasets.

A review of studies clarifies two basic issues. One, among the mentioned studies, there is no study that directly uses similarity methods as a feature selection method. Therefore, as a new proposal, this study directly uses similarity methods as a feature selection method and compares their prediction performance with feature selection methods. Second, in this study, a real data set (Financial data of the 100 largest companies by revenue), to evaluate the sensitivity of each method to the sample size and compare it with another. The rest of the paper is organized as follows: methodology is discussed in Section 2, Section 3 presents the results of the study, and Section 4 reports a discussion of findings and conclusions.

## Methodology

This section elaborates on the methodology adopted for this research work. The complete methodology is depicted in Fig. 1 and consists of the following steps**.**

- Historical finance datasets of the 100 biggest companies are collected
- In this step, appropriate features are selected using feature selection methods and similarity methods
- feature selection methods were used in 80 steps. Each step reduced the dataset size by 1% until just 20% of the primary dataset
- Linear regression is trained on selected features and forecast 10 days ahead of APPL close price.
- In the last step, Linear regression performance is evaluated through cross-validation techniques and results are documented

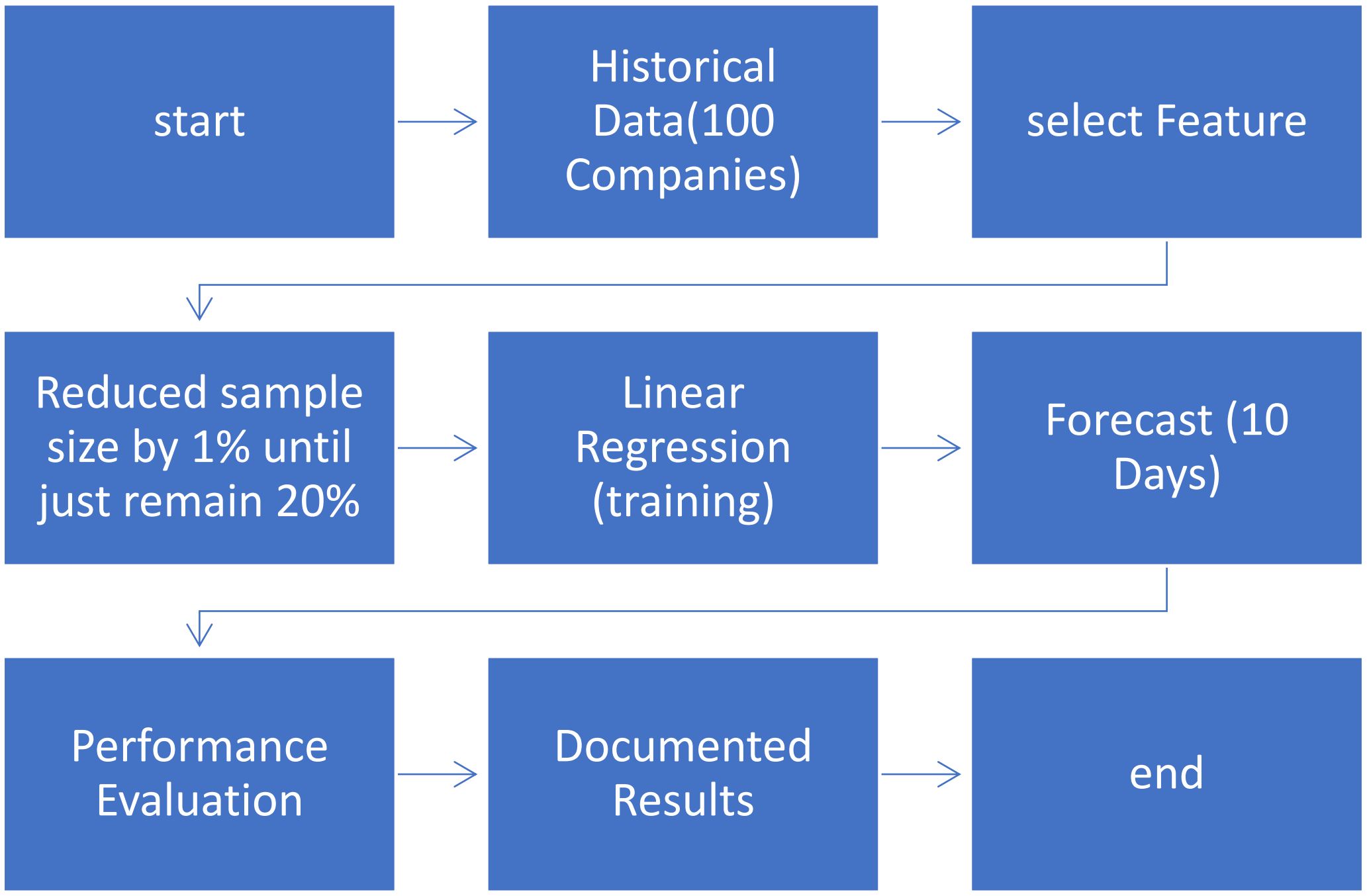

Fig. 1 The complete methodology

**Dataset**

Based on the aim of this paper, to examine the Density and performance of the feature selection methods and similarity methods during high and low sample sizes, the finance dataset was chosen. A large amount of financial data is a suitable feature to examine the performance of methods in large to small amounts of data. According to the Fortune Global 500 2023 rankings, the data set of this research was secondary data including open, low, high, and close prices and the volume of the 100 biggest companies by consolidated revenue. The target value of this dataset was Apple's close price the prediction of the closing price of this variable is done in different data sizes and the best model was selected from among the datasets. The data were collected from the Yahoo Finesse site spanning from January 1, 2016, to January 28, 2024.

This research's main approach is measuring feature selection algorithms' sensitivity to sample size. For this purpose, the feature selection methods were used in 80 steps. Each step reduced the dataset size by 1% until just 20% of the primary dataset.

**feature selection methods**

Once the database without missing value is obtained, the next step is to apply FS and similarity methods to choose the most relevant variables. Feature selection involves the study of algorithms aimed at reducing the dimensionality of data to enhance machine learning performance. In a dataset with N data samples and M features, feature selection aims to decrease M to M′, where $M' \leq M$. Subset selection entails evaluating a group of features together for their suitability. The general procedure for feature selection comprises four key steps: Subset Generation, Evaluation of Subset, Stopping Criteria, and Result Validation. Subset generation involves a heuristic search, where each state specifies a candidate subset for evaluation within the search space. Two fundamental issues determine the nature of the subset generation process. Firstly, the successor generation determines the search's starting point, which influences its direction. Various methods, such as forward, backward, compound, weighting, and random methods, may be considered to decide the search starting points at each state [13]. Secondly, the search organization is responsible for the feature selection process with a specific strategy, such as sequential, exponential, or random search. Any newly generated subset must be evaluated based on specific criteria. Consequently, numerous evaluation criteria have been proposed in the literature to assess the suitability of candidate feature subsets. These criteria can be categorized into two groups based on their dependency on mining algorithms: independent and dependent criteria [14]. Independent criteria exploit the training data's essential characteristics without employing mining algorithms to evaluate the goodness of a feature set or feature.

Based on the selection strategies and/or criteria, there are three main types of feature selection techniques. wrappers, filters, and embedded methods [15]. Wrappers use a search algorithm to search through the space of possible features and evaluate each subset by running a model on the subset. Wrappers can be computationally expensive and have a risk of overfitting to the model. Filters are similar to wrappers in the search approach, but instead of evaluating against a model, a simpler filter is evaluated. Embedded techniques are embedded in, and specific to, a model. The table below illustrates the most well-known methods in each category.

Table1. feature selection methods

| | **Method Name** | **Definition** | **Disadvantage** |
|---|---|---|---|
| **Filters methods** | Correlation-based | Identifies the strength and direction of the linear relationship between two variables. | Assumes only linear relationships and may miss out on nonlinear associations. |
| | Variance Threshold | Eliminates features with low variance, considering them less informative. | It cannot capture nonlinear relationships and might overlook useful features with low variance. |
| | Information Gain | Measures the effectiveness of a feature in classifying the data, often used in decision trees. | They might struggle with continuous data, and selecting features based on information gain may not be sufficient for complex datasets. |
| **Wrappers methods** | Forward Selection | It is a greedy algorithm that starts with an empty set of features and iteratively adds one feature at a time based on certain criteria, such as improving the performance of the model. | Overfitting is a concern, and it is sensitive to the initial set of features, potentially missing global patterns. Despite limitations, it is practical for simplicity in resource-constrained scenarios, requiring careful consideration of criteria for feature selection. |
| | Backward Elimination | Backward Elimination is a feature selection method that starts with all features and iteratively removes the least significant ones based on a chosen criterion, often improving the model's performance at each step. | One disadvantage of Backward Elimination is that it does not allow for the addition of features in later steps, limiting its ability to reconsider decisions and potentially resulting in a suboptimal feature subset. |
| | Recursive Feature Elimination | Recursive Feature Elimination is a feature selection technique that systematically removes the least important features from the model, typically by recursively training the model and assessing feature importance until the desired number of features is reached. It helps identify the most relevant subset of features for optimal model performance. | One potential disadvantage of RFE is its computational intensity, especially when dealing with a large number of features, as it involves repeatedly training the model and evaluating feature importance. Additionally, it may not perform well in cases where features interact in complex ways or when the relationship between features and the target variable is non-linear. |
| | Stepwise Selection | Stepwise Selection is a feature selection method that involves iteratively adding or removing features from a model based on certain criteria. There are two main types: Forward Selection and Backward Elimination. In Forward Selection, features are added one at a time, while in Backward Elimination, features are removed iteratively. The process continues until a predefined criterion, such as model performance or a significance level, is met. | A potential disadvantage of Stepwise Selection is its sensitivity to the order in which features are added or removed, which can lead to suboptimal subsets. Additionally, the stepwise nature may not consider interactions between features effectively, and the final subset chosen may be influenced by the stopping criterion selected. Careful consideration of criteria and potential overfitting is essential when applying stepwise selection methods. |
| | Genetic Algorithms | Genetic Algorithms are optimization methods inspired by natural selection. They iteratively evolve a population of potential solutions, applying genetic operators like crossover and mutation. Fitness evaluation guides the selection of individuals for the next generation. Despite their effectiveness, GAs can be computationally intensive, and tuning parameters is crucial. | A drawback is their computational complexity, especially for large search spaces, and the challenge of parameter tuning. GAs may not guarantee to find the global optimum and are sensitive to parameter choices and problem characteristics. |
| | Simulated Annealing | Simulated Annealing is a probabilistic optimization algorithm inspired by annealing in metallurgy. It explores solutions by allowing both uphill and downhill movements, preventing it from getting stuck in local optima. However, its effectiveness depends on parameter choices, and convergence rates may vary. | Sensitivity to parameter choices, such as the cooling schedule, and variable convergence rates depending on the problem make Simulated Annealing less efficient for certain optimization tasks. |
| **Embedded methods** | L1 Regularization (Lasso) | Lasso is a regularization method in machine learning that adds a penalty term to the model's cost function, promoting sparsity in the coefficients and performing feature selection. | Sensitivity to the choice of the regularization parameter (λ) is a potential drawback, requiring careful tuning for optimal results. |
| | Tree-based methods | Tree-based methods, like Random Forest and gradient-boosted trees, use decision trees to capture complex patterns in data through recursive feature splits. | Prone to overfitting, especially with deep trees, requires regularization techniques and parameter tuning for better generalization. |
| | Recursive Feature Elimination with Cross-Validation (RFECV) | RFECV combines Recursive Feature Elimination (RFE) and cross-validation to iteratively select an optimal subset of features and evaluate model performance. | The method can be computationally intensive due to multiple cross-validation iterations and may have limitations with non-linear relationships and complex feature interactions. |
| | XGBoost and LightGBM | XGBoost and LightGBM are powerful gradient-boosting frameworks in machine learning. They efficiently build decision tree ensembles, but proper hyperparameter tuning is crucial to prevent overfitting. | Both XGBoost and LightGBM can be sensitive to hyperparameter tuning, and improper settings may lead to overfitting. Careful parameter selection is essential for optimal performance. |

## Similarity methods

As is clear in Table 1 each method of FS has some Limitations and weaknesses. Therefore, the time series similarity methods can be a good choice as feature selection methods. Measuring similarity in time series forms the basis for the clustering and classification of these data, and its task is to measure the distance between two-time series. The similarity in time series plays a vital role in analyzing temporal patterns. Firstly, the similarity between time series has been used as an absolute measure for statistical inference about the relationship between time series from different data sets [16]. In recent years, the increase in data collection has made it possible to create time series data. In the past few years, tasks such as regression, classification, clustering, and segmentation were employed for working with time series. In many cases, these tasks require defining a distance measurement that indicates the level of similarity between time series. Therefore, studying various methods for measuring the distance between time series appears essential and necessary. Among the different types of similarity measurement criteria for time series, they can be divided into three categories: step-by-step measures, distribution-based measures, and geometric methods. Table 2 describes both advantages and disadvantages of similarity methods.

Table 2. Similarity methods

| Method | Advantages | Disadvantages |
|---|---|---|
| **Euclidean distance** | The most straightforward, clearest, and most widely used criteria<br>No need for parameter estimation | The exact length of the two-time series,<br>Lack of local time shift support.<br>Inefficiency with increasing dimensions of the time series.<br>The sensitivity of Euclidean distance to small changes in the time axis |
| **DTW) Dynamic Time Warping(** | The DTW interval performs local scaling for the time dimension and ensures the preservation of the order of the time series samples. Any point of the first time series can be compared with any arbitrary point of the second series, provided that their differences are minimized. One of the advantages of this distance function is its ability to measure the distance between time series with different lengths and support the local time shift. | Being time-consuming<br>Sensitivity to noise, the heavy computational load required to find the optimal time-alignment path, incorrect clustering due to a large number of outliers at the beginning and end of the sequence (some elements may not be comparable where DTW must find all elements match.)<br>limiting the time deviation,<br>The need to calculate some costly Lp norms,<br>Not being metric and<br>Need to pair all elements in a series |
| **LCSS) Longest Common SubSequence(** | It is robust against noise and, in addition to giving more weight to similar parts of the series, provides an intuitive concept between paths. By focusing on the common parts, it gives the correct clustering. It enables more efficient approximate calculations. In this method, unlike the Euclidean distance, the data do not need to be normalized. | The results of time series data mining under LCSS strongly depend on the similarity threshold because the similarity measurement approach in LCSS is a zero-and-one approach. Since there is no information about the data and it is tough to determine the correct similarity threshold, using LCSS can lead to poor results. It is not metric and does not obey the triangle inequality. |
| **EDR) Edit Distance on Real sequence(** | Existing distance functions are usually sensitive to noise, change, and data scaling, which usually occurs due to sensor failure, errors in detection techniques, disturbance signals, and different sampling rates. It is not always possible to wipe the data to remove these items. EDR is robust against data corruption. | not metric<br>The reason that it does not obey the triangle inequality is that when a gap is created, it repeats the previous element. |
| **ERP) Edit Distance with Real Penalty(** | ERP is the only distance metric regardless of the Lp norm used, but it works better for regular series, especially for determining the gap g value.<br>ERP is a method based on editing distance that benefits from the advantages of DTW and EDR. This criterion uses a reference point to measure the time gap between matching samples. ERP transforms EDR into a metric whose distance function follows the triangle inequality law. | Because this method involves time thresholding, two locations will not be compared if the difference between their time indexes is too large. |
| **Hasdorf** | Hausdorff distance is a metric measure. It measures the distance between two sets of metric spaces.<br>It shows the spatial similarity between two routes and measures how far they are from each other. | When two curves have a small Hausdorff distance but are not generally similar, in this case, the Hausdorff distance is not suitable. The reason for this disagreement is that the Hausdorff distance only considers the set of points of both curves and does not reflect the trend of the curves. However, the trend is vital in many applications, such as handwriting recognition.<br>Hausdorff distance, in addition to route samples, all points in between<br>It also considers samples, which complicates the calculation of this<br>becomes the standard.<br>They have been widely used in many domains where shape comparison is needed, but they cannot generally compare paths. The Frechet and Hausdorff distance returns the maximum distance between two objects at given points in the two objects. |
| **Fraishe is discrete** | It considers samples and their order in a continuous sequence.<br>Iter et al. used the discrete Freiche distance based on the regression model to estimate the continuous Freiche distance<br>developed This method reduces the complexity of Freishe's | They have been widely used in many domains where shape comparison is needed, but they cannot generally compare paths. The Frechet and Hausdorff distance returns the maximum distance between two objects at |
| | criterion | given points in the two objects. |
| **SSPD )Symmetric Segment Path Distance(** | Because the SSPD method is the sum of the Euclidean distances and considering that it is based on point to segment-, it has solved the problem of the Euclidean method, and it is symmetrical.<br>Like Hausdorff's method, it depends on the distance of the point from the line segment. It calculates the distance of the point from the line segment for all samples of the reference path and all other line segments.<br>Hausdorff uses the maximum point-to-path distance, and SSPD uses the mean, which explains why they have almost the same computation time.<br>This distance is not time-sensitive and compares the shape and physical distance between two path objects. This method does not require any additional parameters or mapping of different routes. | |

Source: [17]

**Measure the performance of methods**

After selecting the most relevant variables and creating a suitable subset, the performance of each of the selected subsets was measured using A 10-fold cross-validation method. This method divided the one dataset randomly into 10 parts. The 9 parts out of 10 parts are used for training and reserved one-tenth for testing[18]. This process was repeated 10 times, reserving a different tenth for testing. During this process, the linear regression model is used for training and testing.

In statistics, linear regression is a statistical model that estimates the linear relationship between a scalar response and one or more explanatory variables. The case of one explanatory variable is called simple linear regression; for more than one, the process is called multiple linear regression. This term is distinct from multivariate linear regression, where multiple correlated dependent variables are predicted, rather than a single scalar variable. If the explanatory variables are measured with error, then errors-in-variables models are required, also known as measurement error models. a linear regression model assumes that the relationship between the dependent variable y and the vector of regressors x is linear.

$$Y = X\beta + \varepsilon \qquad (1)$$

This relationship is modeled through a disturbance term or error variable ε — an unobserved random variable that adds "noise" to the linear relationship between the dependent variable and regressors [19]. Linear regression identifies the equation that produces the smallest difference between all the observed values and their fitted values. To be precise, linear regression finds the smallest sum of squared residuals that is possible for the dataset.

The evaluation metrics used to appraise the performance of the regression models consisted of the coefficient of determination ($R^2$). R2 evaluates the efficiency of feature selection algorithms. The coefficient of determination measures the proportion of the variance in the dependent variable that is predictable from the independent variables. The equation for $R^2$ can be described as follows:

$$R^2 = \frac{Explained\ variations}{Total\ variations} \qquad (2)$$

R-squared evaluates the scatter of the data points around the fitted regression line. It is also called the coefficient of multiple determination for multiple regression. For the same data set, higher R-squared values represent smaller differences between the observed data and the fitted values.

The approach of this article is to identify methods of feature selection and similarity that have the best performance in small data sizes. For the discussion of feature selection, in addition to the methods proposed in the research method, there are many methods in the research literature, including the combined methods that are more accurate in many cases. However, the main aim of this study is to identify methods that are computationally simple in addition to selecting the most suitable subset of data. Therefore, methods were chosen that did not have computational complexity.

**Result**

The performance of feature selection methods in different data sizes was evaluated. The average value of r-squared was measured in each step of reducing the sample size and selecting an appropriate data subset based on existing methods (Table 3). The Var methods have the best performance based on the values of r-squared. The stepwise and correlation took the next position. The lasso method has the worst performance between other methods. Among the 15 feature selection methods used in this study, the Euclidean distance method ranks 6th and the DTW method ranks 7th, which performed better than other similarity methods. The edit distance similarity method has the worst performance.

Table3. The average value of r-squared

| **methods** | **r-squared** |
|---|---|
| var | 0.996092 |
| stepwise | 0.995537 |
| cor | 0.994809 |
| backward | 0.99406 |
| Recursive Feature Elimination | 0.993215 |
| EU | 0.991367 |
| dtw | 0.991285 |
| simulated | 0.990358 |
| tree-base | 0.9879 |
| forward | 0.980137 |
| Hausdorff | 0.977381 |
| mutual information | 0.977135 |
| frechet | 0.976615 |
| edit distance | 0.951547 |
| lasso | 0.704437 |

Figure 2 shows the r-squared value of different feature selection methods in different datasets. In fact, by reducing the sample size in each step, different data sets were selected according to different methods. The horizontal axis of the graph represents the remaining percentage of the number of observations. In each step, by reducing the number of samples and choosing one, considering that the number of observations has decreased, the performance of the regression model decreases and the r-squared value in the number of low samples is lower than the number of high hub samples. In general, the value of r-squared was low in all methods at low percentages, and as the percentage increased, its value increased, and the performance of the 15 existing methods was similar and close to each other. However, the r-squared value of the lasso method was dramatically lower than other methods.

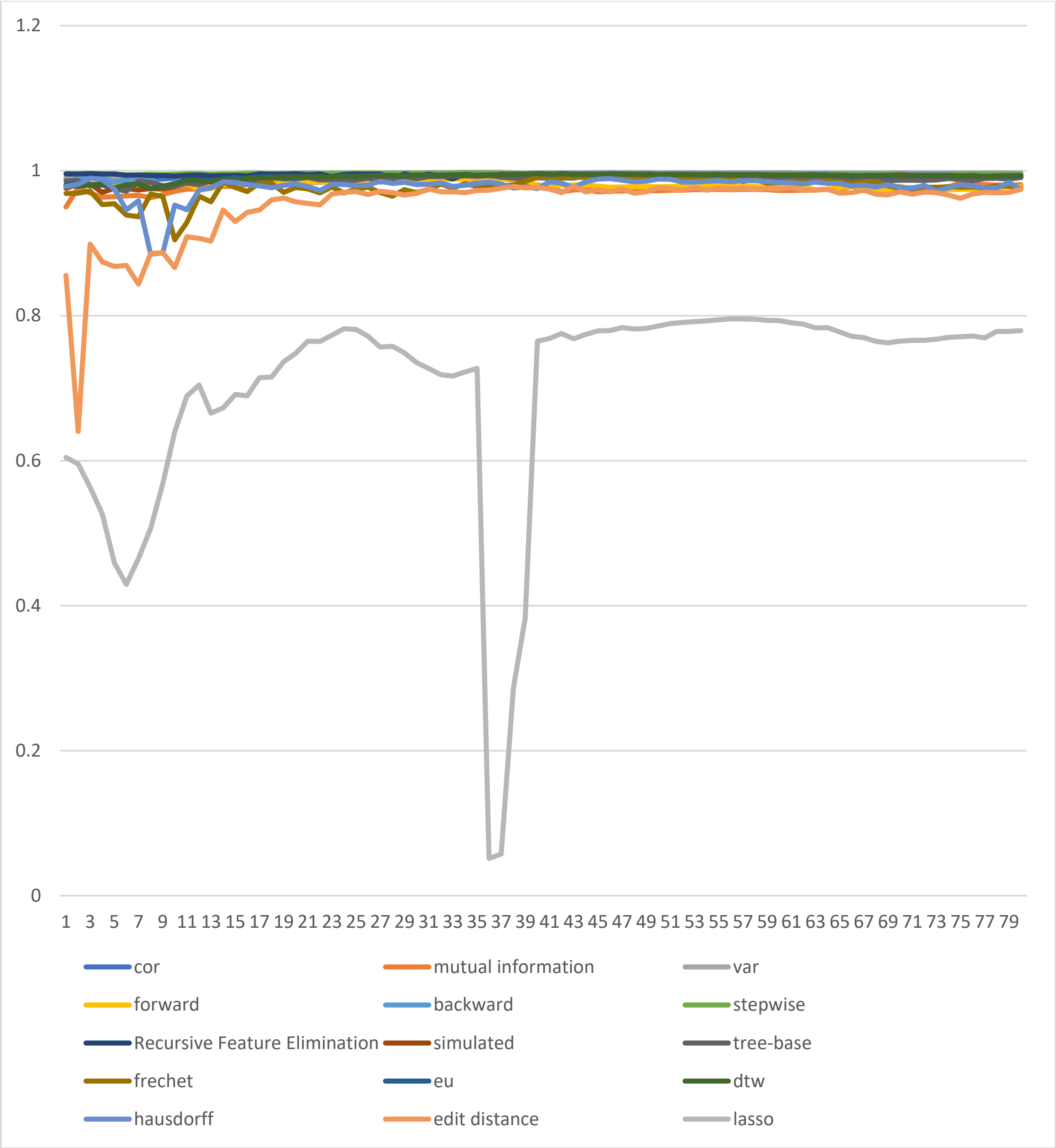

Fig2. The value of r-squared of feature selection methods in the number of different observations

The figure shows the r-squared value of filtered methods. The trend line of each of these graphs was drawn. Regarding the slope of the trend line, among the three existing methods, the mutual information method had the lowest slope and sensitivity to the number of observations. However, the fluctuations of the r-squared value were high, which made this method less reliable. On the other hand, although the slope of the trend line of the var method was slightly higher than the mutual information method, the r-squared changes during the change in the number of observations were less than the other methods of this group.

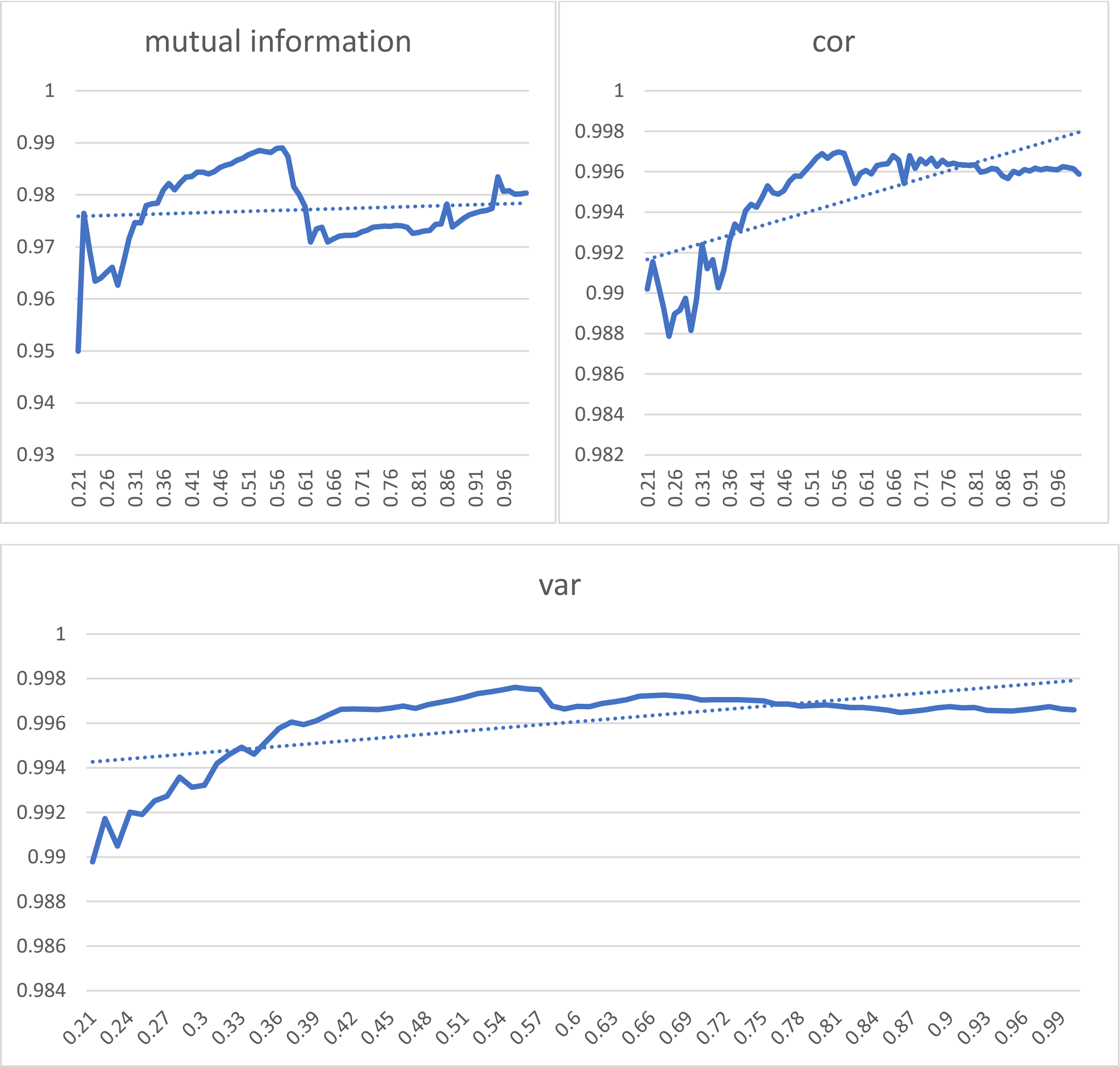

Fig3. The value of r-squared of filtered feature selection methods

The performance results of the Wrappers methods are shown in Figure 4. Five known methods from this group were reviewed. Among these, the value of r-squared fluctuated greatly during the change of the number of samples in forward, recursive feature elimination, and stepwise methods. Among the two backward and simulated methods, the simulated method had less fluctuation with a lower slope.

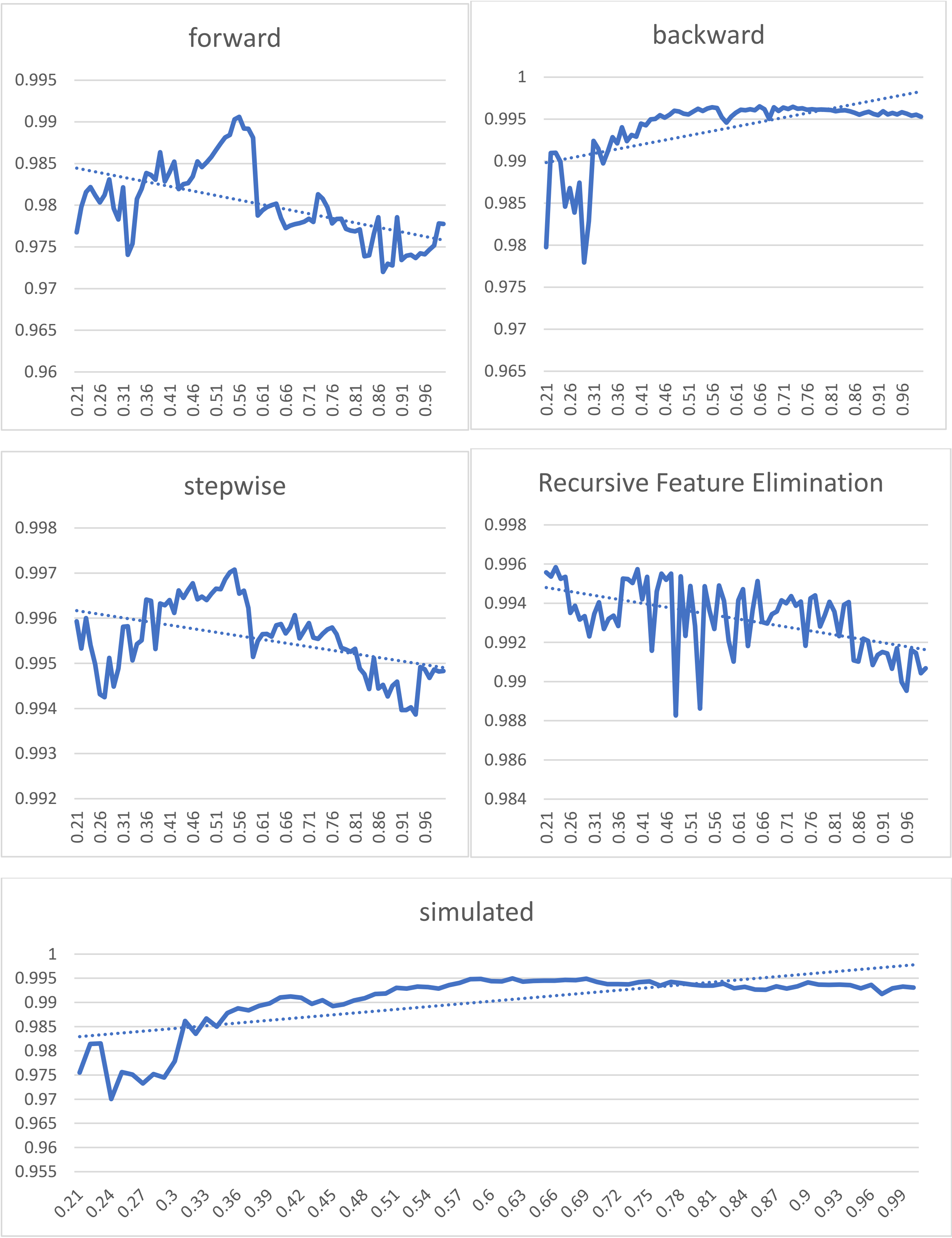

Fig4. The value of r-squared of Wrappers methods

From the group of embedded methods, two methods were investigated (Fig5). The Lasso method had a relatively lower slope than Tree-based. However, the r-squared value of this method is lower than the tree-based method.

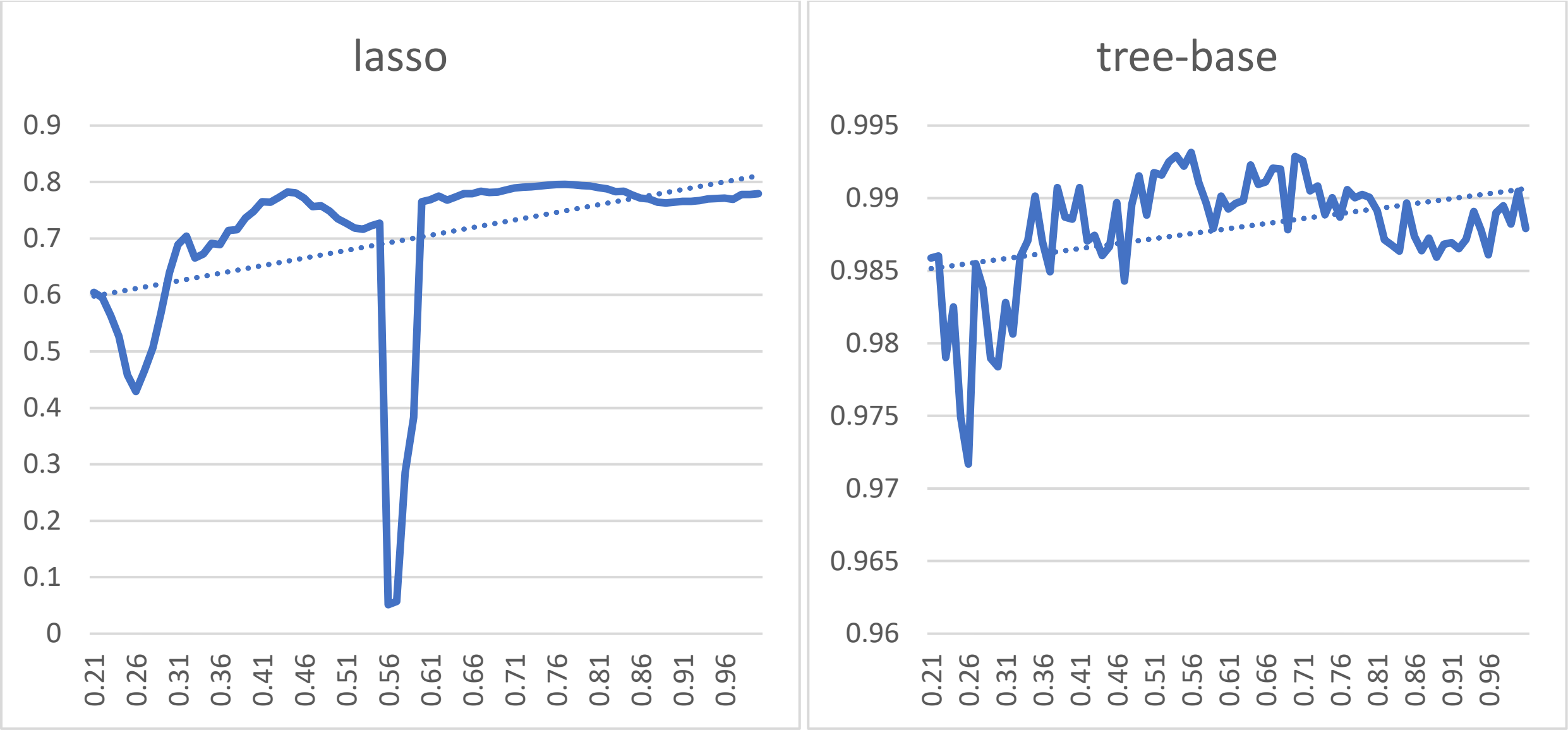

Fig5. The value of r-squared of embedded methods

Figure 6 shows the performance of 5 similarity methods. among these five methods, the edit distance method had the lowest slope. Similarity methods had minor fluctuations during data size reduction.

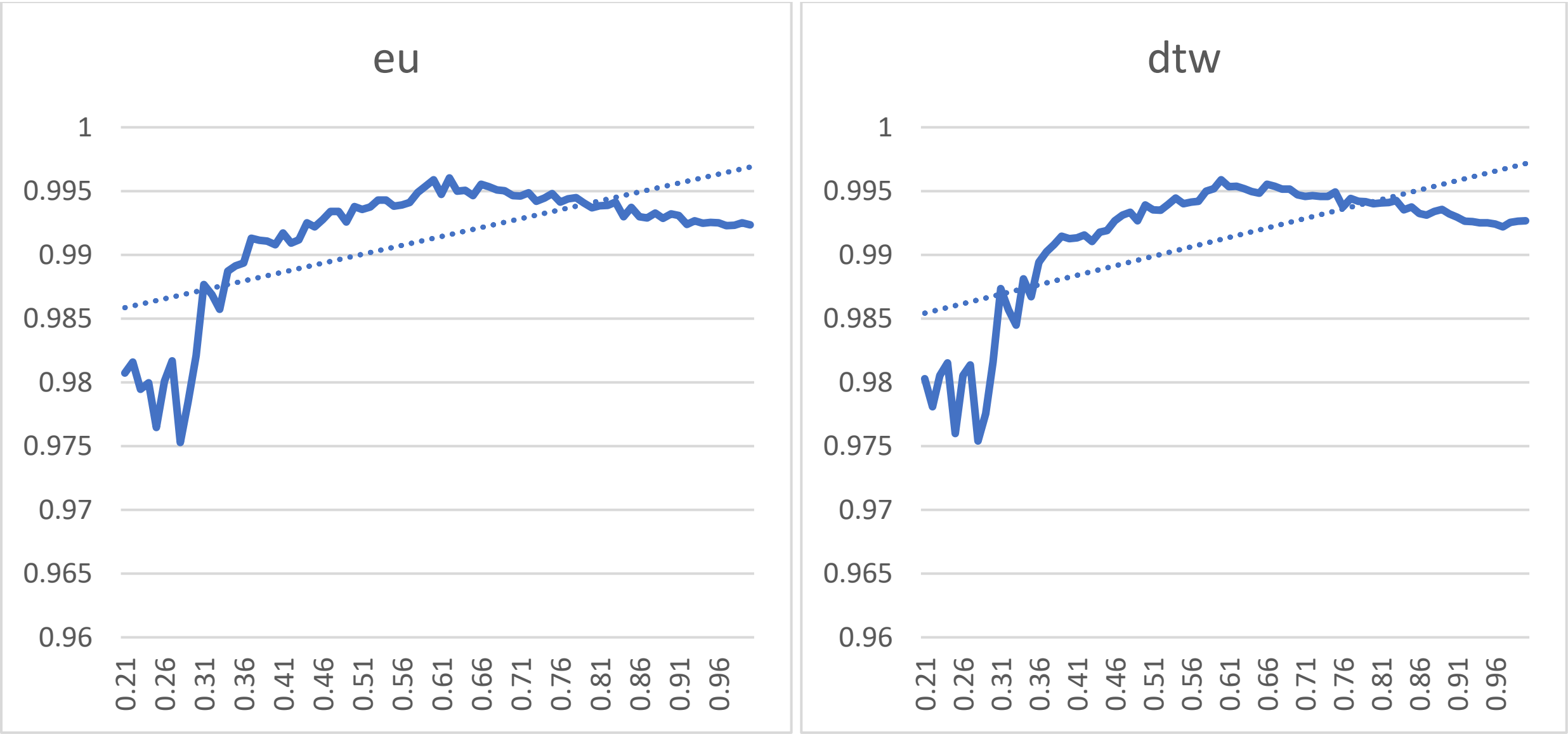

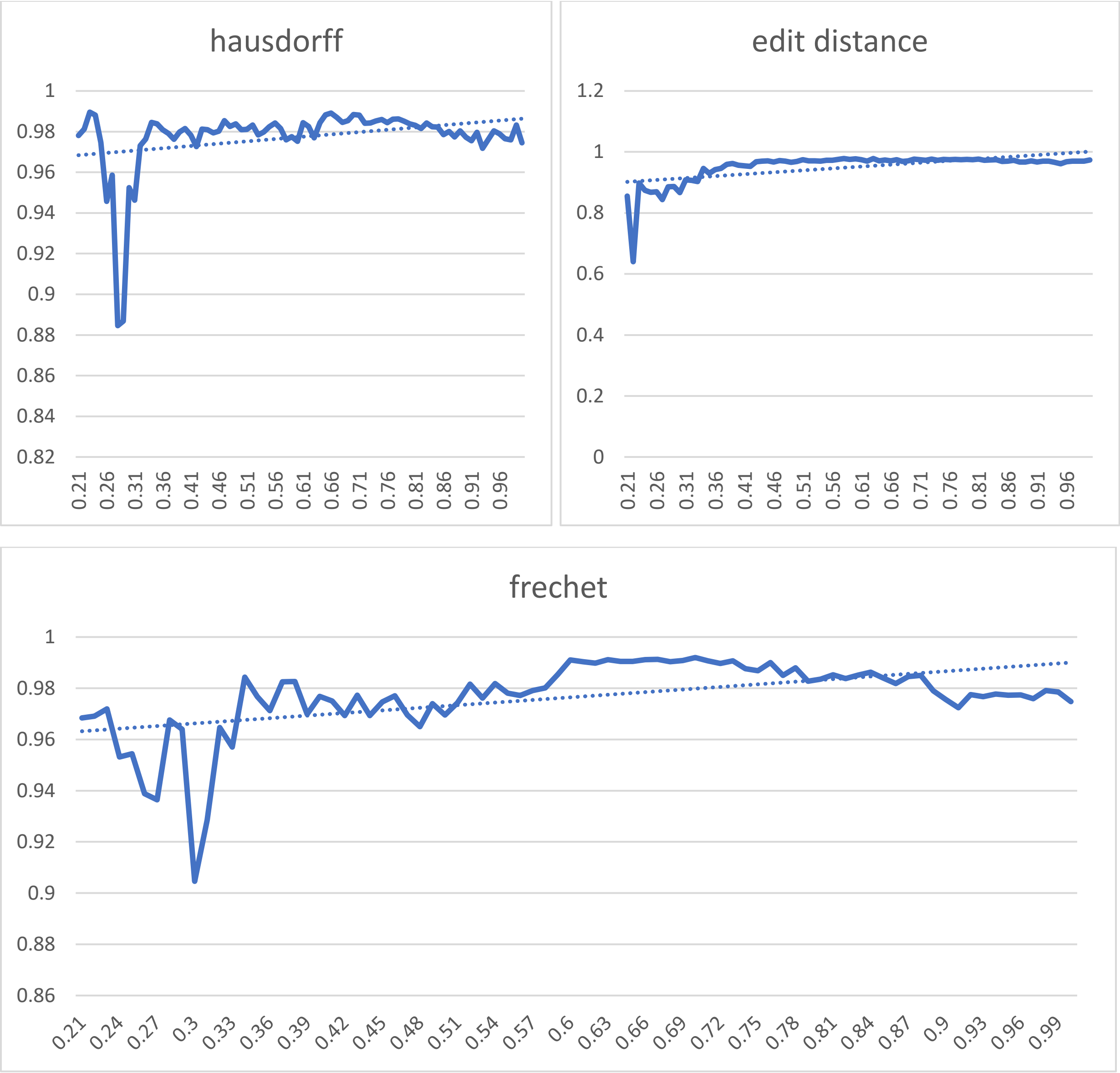

Fig6. The value of r-squared of similarity methods

**Discussion**

There are three main factors to choose the best method in this research; the value of r-squared, the sensitivity of methods to change in sample size, and fluctuation during change in sample size. Among the methods, on average, variance, stepwise, and correlation methods had the highest r-squared value. The mutual information, var, simulated, edit distance, and Hausdorff methods had less sensitivity to data size. The Variance, simulated, and edit distance had less fluctuation. According to these metrics, the var method had all three criteria. Among the similar methods, Hausdorff and edit distance had good performance in comparison to other methods.

this study focuses on improving the performance of the data-driven model by selecting the most appropriate features. However, it is important to acknowledge certain limitations of this study. First, many existing hybrid methods should be investigated. Second, the results presented may change as the dataset changes.

**Conclusion**

The aim of this study is to select feature selection techniques that have little sensitivity to low data size and select a subset of data that has high predictive performance in low data size. As mentioned in the results section, based on the dataset used in this study, the performance of standard feature selection methods fluctuates in different data volumes, which can reduce the level of confidence in these techniques. Among the ten standard feature selection methods, two variance and simulated methods are more stable than others. The graphs show that the similarity methods introduced as an alternative to the feature selection methods are less volatile than these methods, which increases the confidence in the results of these methods when the data size is different. Therefore, according to the first approach of this study, the similarity methods are more reliable than the usual feature selection methods. Of course, it is essential to note that these results are only related to one data set, and the results may change in other data sets.

The second and more critical approach that the article sought to test is to examine the sensitivity of the methods to the change in data size. Indeed, any method with the least minor sensitivity to data size change will be chosen. According to trendline results, the variance, correlation methods, simulated methods, edit distance, and Hausdorff are less sensitive to observation size. Considering that time series similarity methods had the most minor fluctuation among other feature selection methods, these methods can be used as reliable methods for feature selection. Similarity methods, such as the Hausdorff and edit distance approaches, emerged as the most stable among the various feature selection techniques evaluated. This robustness across different data sizes underscores their reliability and suitability for this research context. Their consistent performance indicates that they can effectively handle fluctuations in observation numbers without significant loss of predictive accuracy. This resilience is crucial in ensuring the robustness and generalizability of predictive models, particularly in dynamic environments such as financial markets. Consequently, these methods stand out as promising tools for feature selection, offering researchers a dependable approach to identifying relevant variables for predictive modeling tasks, such as forecasting Apple's closing price.